\documentclass[conference]{IEEEtran}
\IEEEoverridecommandlockouts
\usepackage{amsmath,amssymb,amsfonts}
\usepackage{subcaption}
\usepackage{multirow}
\usepackage[table,xcdraw]{xcolor}
\usepackage{algorithmic}
\usepackage{graphicx}
\usepackage{textcomp}
\usepackage{xcolor}
\usepackage{hyperref}
\def\BibTeX{{\rm B\kern-.05em{\sc i\kern-.025em b}\kern-.08em
    T\kern-.1667em\lower.7ex\hbox{E}\kern-.125emX}}

\begin{document}

\title{Evaluation of MQTT Bridge Architectures in a Cross-Organizational Context
\thanks{This work was supported by SFI SmartOcean NFR Project 309612/F40.}
}

\author{\IEEEauthorblockN{Keila Lima\IEEEauthorrefmark{1}, Tosin Daniel Oyetoyan\IEEEauthorrefmark{1}, Rogardt Heldal\IEEEauthorrefmark{1}, Wilhelm Hasselbring}\IEEEauthorrefmark{2}
\IEEEauthorrefmark{1}Western Norway University of Applied Sciences, Norway \\
Email: \{keila.lima, rogardt.heldal, tosin.daniel.oyetoyan\}@hvl.no \\
\IEEEauthorrefmark{2}Kiel University, Kiel, Germany \\
Email: hasselbring@email.uni-kiel.de
}

\maketitle

\begin{abstract}
The latest surveys estimate an increasing number of connected Internet-of-Things (IoT) devices (around 16 billion) despite the sector's shortage of manufacturers. All these devices deployed into the wild will collect data to guide decision-making that can be made automatically by other systems, humans, or hybrid approaches. 
In this work, we conduct an initial investigation of benchmark configuration options for IoT Platforms that process data ingested by such devices in real-time using the MQTT protocol. We identified metrics and related MQTT configurable parameters in the system's component deployment for an MQTT bridge architecture. For this purpose, we benchmark a real-world IoT platform's operational data flow design to monitor the surrounding environment remotely. We consider the MQTT broker solution and the system's real-time ingestion and bridge processing portion of the platform to be the system under test. In the benchmark, we investigate two architectural deployment options for the bridge component to gain insights into the latency and reliability of MQTT bridge deployments in which data is provided in a cross-organizational context.
Our results indicate that the number of bridge components, MQTT packet sizes, and the topic name can impact the quality attributes in IoT architectures using MQTT protocol. 
\end{abstract}


\section{Introduction}
\label{sec:intro}
The 2023 report on the IoT market predicted an 18$\%$ increase in the number of deployed devices in the previous year~\cite{IoTAnalytics2023}. The same report estimates a continued increase in this number by at least 16$\%$ until 2025 which will result in more than 16 billion devices if the prediction is fulfilled.
Many of these deployed systems are already delivering data as a service via sensor data provider organizations to other organizations that use them in their daily operations. This interfacing raises interoperability issues when integrating with the sensor provider’s systems~\cite{aziz21}. In this context, business models for IoT data marketplaces emerge where data variety is pointed out as one of the biggest challenge of such systems in order to provide ``\textit{value-added services}'' to data consumers~\cite{8664564}. The Message Queuing Telemetry Transport (MQTT) standard appears as an adequate application-level communication protocol in these IoT contexts helping to bridge the sensor provider systems with the data consumer systems.

\begin{figure}[htb!] 
	\centering 
	\begin{subfigure}{\columnwidth}
    	\centering 
    	\includegraphics[width=0.6\textwidth,trim={0cm 4.5cm 13cm 3.5cm},clip]{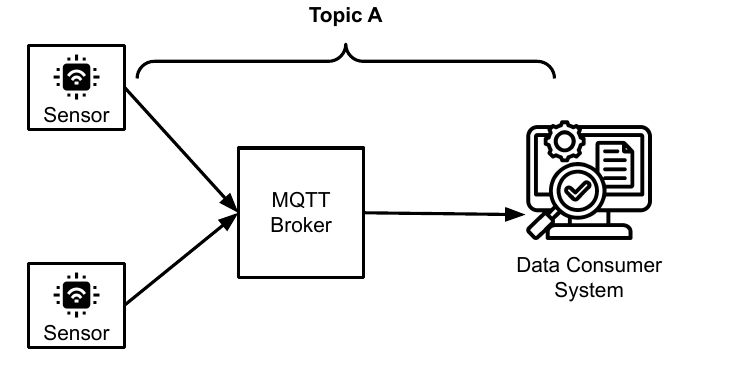} 
    	\caption{Standard MQTT broker (single instance/cluster) deployment.} 
    	\label{fig:mqtt_std}
	\end{subfigure}
	\hfill
	\begin{subfigure}{\columnwidth}
    	\centering 
	\includegraphics[width=0.9\textwidth,trim={3.8cm 3cm 3cm 2.5cm},clip]{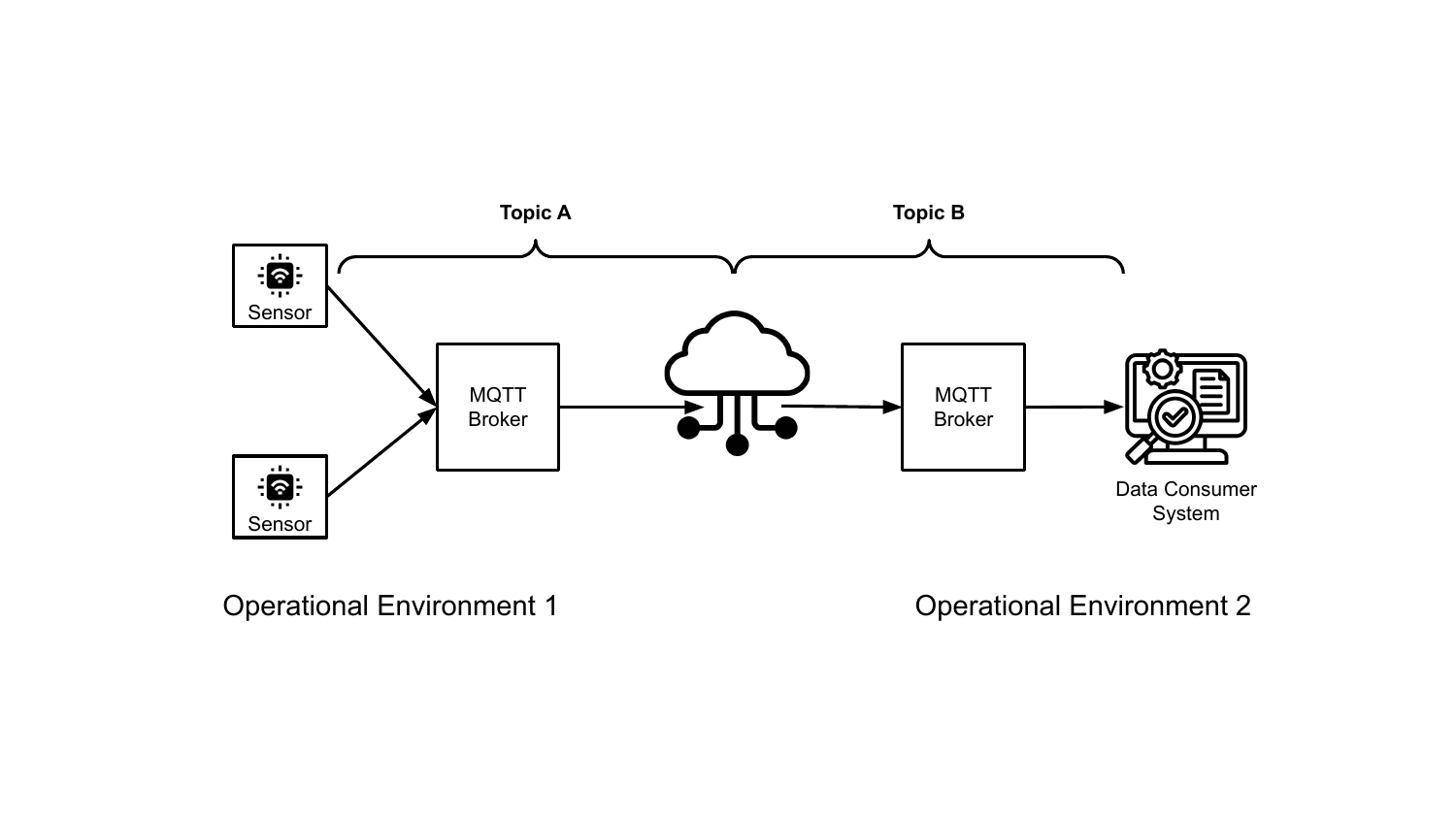} 
	\caption{MQTT broker bridge deployment. }
	\label{fig:mqtt_bridge}
	\end{subfigure}
	\caption{MQTT example communication patterns.}
    \vspace{-0.5cm}
	\label{fig:mqtt}
\end{figure}

MQTT is a publish-subscribe architecture where publishers (e.g., IoT sensor processes) publish on a topic (e.g., Topic A) configured in a MQTT broker and values are pushed by the broker to subscribers (e.g., Data Consumer System) that subscribed to the same topic (see Figure~\ref{fig:mqtt_std}). This simple architecture does not apply to real-world IoT data scenarios where producers and consumers are often separated by location and network. As a result, the huge data volume that is generated in these remote locations have to be transferred over communication links to data platforms for access by consumer applications. A bridge architecture is thus an approach to push data from a producer broker to a consumer broker that sits in a different network (see Figure ~\ref{fig:mqtt_bridge}). 
Different use cases arise for a bridge architecture in practice. For example, aggregating MQTT topics of the publishers can be an efficient strategy to reduce the amount of traffic during publisher broker to subscriber broker communication~\cite{Soua2022}. Furthermore, in-stream sensor data processing requirements can arise from performing quality analysis in the middleware layer~\cite{liu2020data}. For example, semantic analysis that compares the validity of data using different IoT sensor parameters or the relationship between observations around a specific location. Other examples include data filtering and preparation (e.g., anonymizing) for sharing across domains or use cases~\cite{PETRAKIS2018156}.  

There are different ways that bridge architecture can be implemented in MQTT. A common approach is the implementation on the broker-side~\cite{Longo2022,Soua2022} as provided by the Eclipse Mosquitto. This approach is feasible when there is no administrative barriers, no requirement for data transformation, and the MQTT solution is homogeneous. Current MQTT broker vendors can offer both data transformation and or processing using rules or policies, and bridging as separated services. However, in real-world scenarios, data can be provided and published by different providers who are using different MQTT solutions and are in heterogeneous formats thus creating a requirement for data transformation and a requirement to harmonize them for analytic applications before forwarding to another MQTT broker~\cite{Lima2024}.

In this IoT sensor data-sharing paradigm, additional challenges are faced when data streams are merged in a cross-organizational setting~\cite{Lima2024}. There are administrative barriers and trust boundaries that need to be addressed in the integration of the data producer and data consumer software architectures. 

To address these challenges in a cross-organizational settings, some studies have proposed a client-side bridging architecture for handling heterogeneous data stream together with a data transformer function between MQTT brokers ~\cite{Lima2024}. In this study, we evaluate the latency and reliability of deploying two variants of such architecture when implemented on the client-side of the broker. In addition, we evaluate the overhead of topic name and the payload size and their effects on latency and reliability of the architectures. To infer the identified quality attributes, we have used metrics measured in the system: the number of lost messages for reliability and the end-to-end delay in milliseconds (ms) for latency. 

We have used a benchmark methodology to investigate the architectural deployment options for MQTT bridge solutions implemented on the client side. Performance benchmarks are part of the measurement-based approaches in the field of Software Performance Engineering~\cite{smith2002performance}. In empirical software engineering, benchmarks can be used for comparing different methods, techniques and tools~\cite{EASE2021}. 

The overall contribution of this study is twofold:
\begin{enumerate}
    \item We show benchmark results for MQTT bridge architectures with lower throughput but latency and reliability concerns for real-world IoT systems that stakeholders in practice can consider before designing their bridge architectures. 
    \item We demonstrate that additional benchmark parameters for MQTT-based systems under test, such as data size and topic name, should be considered. These parameters have implications for the latency and reliability of the MQTT bridge architecture.
\end{enumerate}

The rest of the paper is structured as follows: In Section~\ref{sec:background}, we introduce the role of MQTT in IoT architectures. Section~\ref{sec:method} explains our benchmarking methodology, before the experiment results which are presented in Section~\ref{sec:results}. In Section~\ref{sec:threats}, we discuss potential threats to validity of the study. Finally, we compare our approach to other studies in Section~\ref{sec:related-work}, and conclude the paper in Section~\ref{sec:conclusion}.
\section{MQTT in IoT Architectures}
\label{sec:background}
IoT architecture in some approaches, has been defined by five main layers~\cite{6424332}: the perception layer, network layer, processing layer, application layer, and the business layer.
The perception layer is the one responsible for the devices sensing and actuation, interacting directly with the surrounding physical environment. Secondly, the network layer is responsible for the channeling of data to/from the sensing (perception) layer to the processing layer, typically at the cloud side or an infrastructure with access to more computational resources. The processing layer manages data flows performing operations such as aggregations, analysis and storage, and serving data to the domain-specific IoT application where the application layer resides. The last layer, the business layer, is responsible for managing the overall IoT system, including building business models~\cite{6424332,munccini18}.
For the network layer, there is no universal or best-fit option for the communication protocol at the application level, but MQTT is often employed because of its lightweight footprint, having a header of 2 bytes, QoS levels, ordered message delivery, and less complexity characteristics~\cite{mishra2020use}.  
In a systematic mapping study in~\cite{munccini18}, up to six layers were identified, adding an adaptation layer that is responsible for handling device interoperability inside the sensor network. 

In a relative comparison done in~\cite{naik17}, MQTT appears as the best good option when considering the latency and reliability performance.
MQTT is a standard application level protocol managed by OASIS, first created in 1999. It was designed to be lightweight in terms of communication overhead (small headers) and reliability (embedded QoS levels) and thus suited for IoT applications~\cite{mqtt}. The protocol follows the publish/subscribe communication model for message exchange between actors. The protocol has two different client actors, publishers and subscribers, that can communicate via the third one: the broker.
The capacity of the broker to handle the clients is constrained by the queue size, memory, and CPU.

The packet structure in MQTT has a header and a payload section that can optional depend on the type of control packet. There are 16 types of control packets, and they are related to the connection, disconnection, publishing, subscribing, and authentication~\cite{mqtt}.
Concerning message exchange, publishing clients provide data via a topic name. This topic name is verified against a topic filter specified by subscribing clients. The name is often structured by levels that follow some business logic hierarchy, such as geographic location and sensor arrangement in the system.
Lastly, the protocol defines three levels of quality-of-service (QoS)~\cite{mqtt}. The first, ``At most once'' (QoS level 0), does not guarantee the delivery of the message to the broker and thus it is prone to message lost. The second, ``At least once'' (QoS level 1), guarantee the delivery of the message to the broker but duplicates can occur. The last one,  ``Exactly once'' (QoS level 2), guarantee the delivery of the message to the broker but requires double the number of message exchange when compared with the QoS level 0.

There are three broker deployment options when using MQTT protocol for the network layer of IoT systems: single broker deployment, cluster of brokers, and bridge of brokers. The basic option is a single broker instance deployment that handles all publisher and subscriber connections, which includes managing information regarding topics and authorization while handling message traffic. In the publish/subscribe model, the broker becomes a bottleneck, because of the limit of allocated resources, queue size, and message storage (persistency) for failure recovery. Other technique used to improve the fault tolerance is to deploy a cluster of brokers, by replicating single instances deployments and adding links among them for synchronization and consistency of the logic information needed to manage MQTT implementations. This architecture brings greater throughput and capacity to handle a large number of concurrent connections, but as a trade-off, the inter-broker communication brings overhead in the latency~\cite{Longo2022}.
Lastly, there is the bridge architecture between brokers that enables forwarding of messages produced in one environment or part of the system to be decoupled from another section.

\subsection{MQTT Bridge}
\label{sec:mqtt_bridge}
In a MQTT bridge architecture there is a data flow between at least two brokers, typically imposed by the need to separate operational fields where sensors sit from the data delivery pipeline inter or cross-organizational.
The rationale behind the bridge of MQTT brokers is that data is published in one topic on the source broker will be made available in a topic to subscribers in the receiving broker. 
In the latest version of the standard, MQTT 5.0~\cite{mqtt}, launched in 2019, there are some non-normative comments referring to features that can be used to implement the bridge architecture. These features are associated to the options when subscribing to topics on the source broker, namely the ``No Local'' and the ``Retain As Published'' flags~\cite{mqtt}. The ``No Local'' flag when publishing is important to avoid creating loops because if triggered, the publisher client will receive its own message. 
Common applications are to separate the sensor network from the data storage, aggregation and analytics (usually at the cloud side), or in industrial IoT settings, to separate different productions lines.
Examples of such scenario that motivates the usage of the broker-to-broker bridge in literature are presented in~\cite{Lima2024,Soua2022,sigpro20}. The first scenario relates to handling of heterogeneous sensor data streams processing.
Another scenario relates to the optimization of a constrained link over satellite for data distribution for subscribers in the receiving side broker. The particularity of this solution is that the pre-processing is performed in the sending broker side, having access to MQTT components in both sides. Lastly, the geographical separation of MQTT clients is the main reason for adopting the bridge architecture. In the last two cases, the system is operated by the same organization, having access to the software components in the processing units where data is collected and published. While in the first, data is being delivered as a service, having no access to components on the publishing side. 
Additional use cases with similar driver, i.e the separation of modules in different deployment infrastructure environments, are also described by MQTT solutions vendors~\cite{emqx_bridge,cedalo_bridge}.

This study focuses on the link between the edge and the cloud in IoT systems, which also requires data format conversion for the data provided by the organizations operating these sensor-based systems.
In Figure~\ref{fig:mqtt_bridge}, we depict an example of such a scenario, considering the data provision context described in Section~\ref{sec:intro}. In this example, the bridge is implemented on the MQTT client-side, and the topic \textit{A} where data is published on the source broker is subscribed by the bridge component that provides transformed data into topic \textit{B} to subscribers in the receiving broker.

\subsection{Security in  MQTT}
\label{sec:sec}
Compared with the existing alternatives, MQTT has fewer features in its specification when securing the communication link~\cite{aziz21}. Although many non-normative recommendations are suggested in the standard, how these are enforced depends on the implementation~\cite{8897627}. Authentication can be done by user and password definition or by using enhanced methods that require additional message exchange when clients establish a connection to a broker~\cite{mqtt}. Authorization is possible with access control level (ACL) definitions in some implementations (e.g., with dynamic security in Mosquitto). Furthermore, there is support for encryption of the communication channel using TLS/SSL. Alternative authentication and authorization can be implemented around the infrastructure where clients and brokers operate but lay outside the standard specification. Lastly, there are also non-normative recommendations for security measures and options in the standard. One example is the encryption of the payload for message integrity beyond TLS/SSL in devices that can accommodate the different existing algorithms~\cite{mqtt}.

\section{Methodology}
\label{sec:method}
In this section, we describe our context, research questions, the architectural deployment options under test, the load profile and the origin of the replicated data, and the MQTT-related parameters used in the different benchmark setups.
\begin{figure*}[htb] 
	\centering 
	\includegraphics[trim={1.5cm 0.3cm 0.5cm 0cm},clip,width=0.7\textwidth]{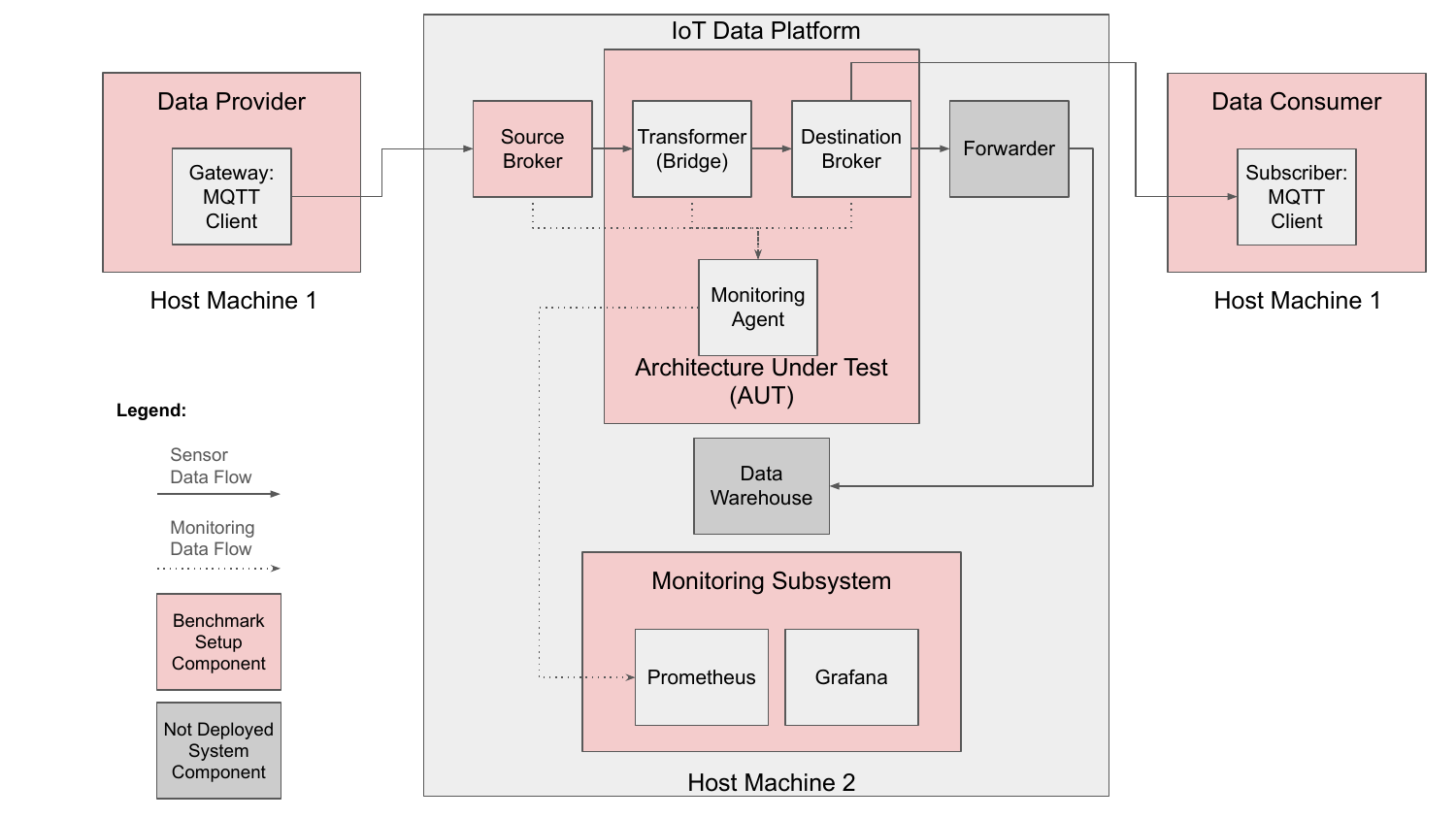}
	\caption{Benchmark setup components in the overall data delivery flow architecture.}
	\label{fig:case}
\end{figure*}

The MQTT bridge architectures we evaluate in this study have the following characteristics:
\begin{enumerate}
    \item A gateway node serves as a sink of data at the edge of operational fields where sensors are deployed, concentrating higher volumes of data instead of sensors transmitting data individually.
    \item Real-time systems require data for decision-making as soon as transmitted and require stream processing on ingested data from gateways. Such systems require less delay (low latency). 
    \item Reliability of message delivery associated with the (over-the-air) transmission of data from the gateways at the edge and cloud hosted systems.
\end{enumerate}
These architectures introduce conflicting quality attributes in the end-to-end data flow, i.e.
Latency \textit{vs.} Reliability. This trade-off is also present in the MQTT protocol, captured by the
quality-of-service (QoS) levels defined in the standard. Since more reliable levels (1 and 2) require additional messages exchange, it results in the increase of the end-to-end delay.

\subsection{Research Questions}
\label{sec:rqs}
The study's overall goal is to evaluate latency and reliability deployment options for the MQTT bridge architecture when implemented with processing capabilities on the client side, and when sensor data is provided in a cross-organizational context.
The quality attributes selected as targets for the benchmark are the same as the IoT platform use case, which aims to monitor operational fields remotely. In this case, how fast data arrives at the application layer for decision-making, and also the availability of data.
The end-to-end latency was measured by finding the difference between the time message was published by the gateway and the time the message was received by the subscriber (see Figure~\ref{fig:case}). The end-to-end message loss was calculated by finding the difference between the number of published messages by each gateway and the number of messages that arrived at the subscriber end from that specific gateway.
We aim to understand which deployment options are more latency-efficient, i.e., allow faster data delivery to the application layer, and and reliability-efficient, i.e., which parameters make the data delivery more reliable.
Therefore, we investigate the following research questions:
\newline
\noindent
\emph{\textbf{RQ1}: How does the deployment architecture of MQTT bridge components impacts message delivery latency and reliability in a cross-organizational data delivery context?}
~\newline
\noindent
We aim to investigate the latency and reliability of different bridge deployment options which are two instances of the same higher-level architecture presented in Figure~\ref{fig:case}.
~\newline
\noindent
\emph{\textbf{RQ2}: How does the topic name and the payload size impact both latency and reliability of the deployment architecture of MQTT bridge components?}
~\newline
\noindent 
We intend to investigate the overhead of the topic name and the payload size, which are the sections that occupy most of the size of MQTT packets. Our aim is to understand how those parameters and the software components using them should be configured when data delivery latency and reliability are of concern.
To answer this research question we rely on the sizes of the different topic names configured for AUT~1 to be able to compare based on the same number of links and deployed bridge processor components.

\subsection{Benchmark Setup}
\label{sec:setup}
\begin{figure*}[htb!] 
	\centering 
	\begin{subfigure}{0.7\textwidth}
    	\centering 
    	\includegraphics[width=\textwidth,trim={0cm 0cm 0cm 0cm},clip]{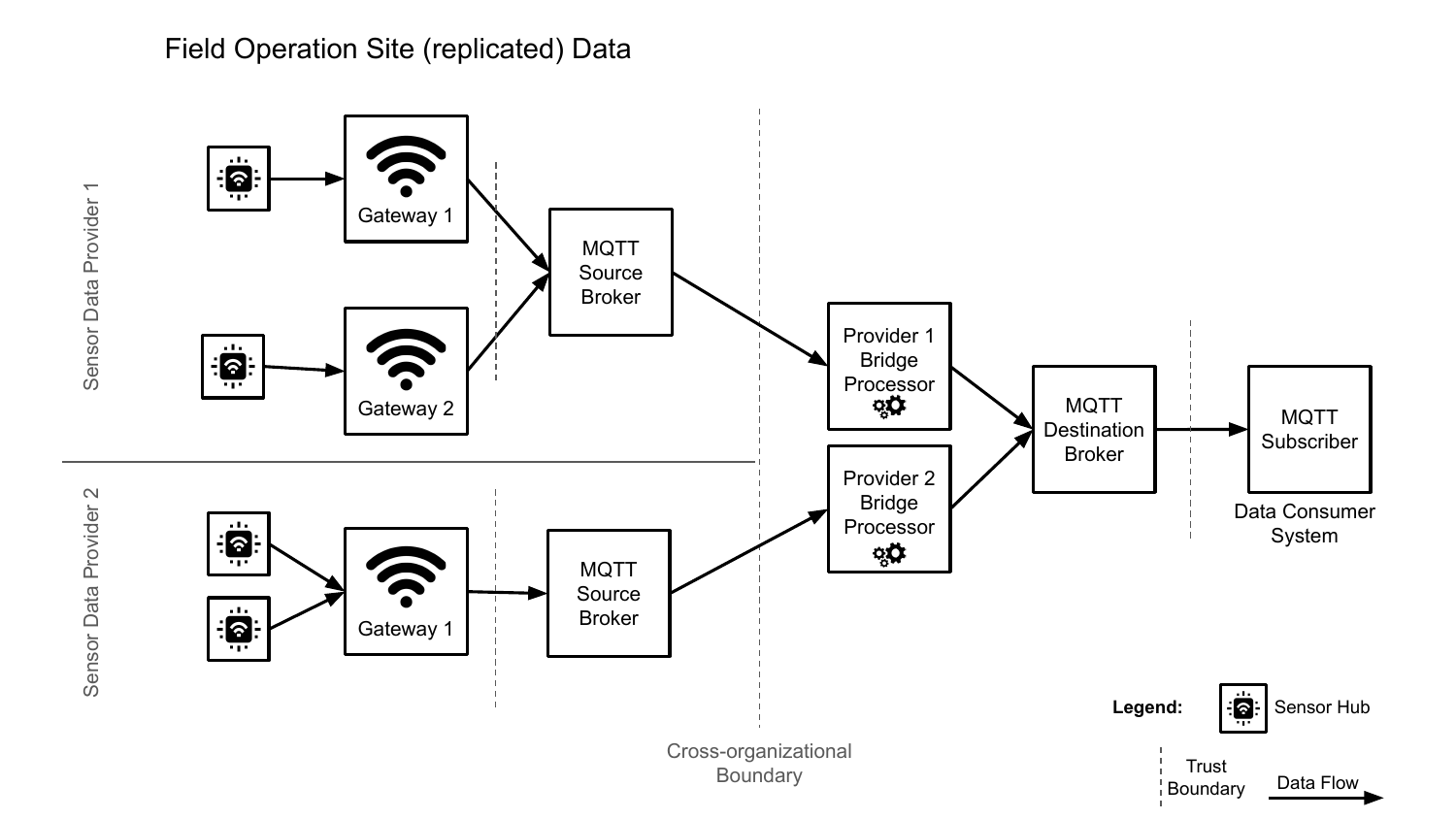} 
    	\caption{Architectural deployment option 1 (AUT 1): One bridge component per sensor data provider.} 
    	\label{fig:aut1}
	\end{subfigure}
	\hfill
	\begin{subfigure}{0.7\textwidth}
    	\centering 
	\includegraphics[width=\textwidth,trim={0cm 0cm 0cm 0cm},clip]{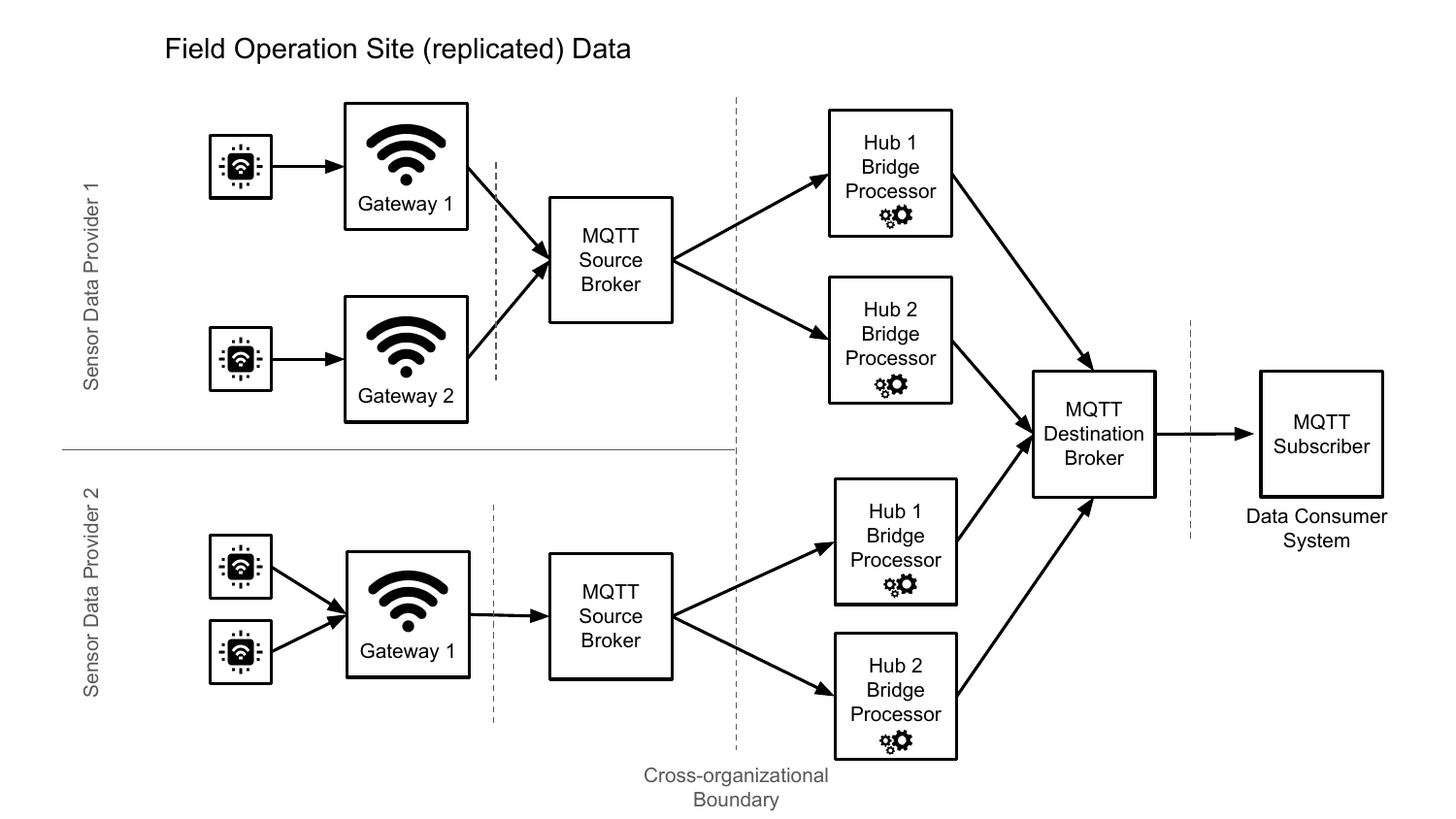} 
	\caption{Architectural deployment option 2 (AUT 2): One bridge component per sensor data stream.}
	\label{fig:aut2}
	\end{subfigure}
	\caption{Architectural Options Under Test (AUT).}
	\label{fig:aut}
\end{figure*}

We employ a benchmark methodology to collect quantitative data in an automated manner of MQTT bridge components from a software prototype of a real IoT system ~\cite{Lima2024}.
Figure~\ref{fig:case} shows the benchmarked IoT Data Platform prototype, while Figure~\ref{fig:aut} shows two deployment architecture options of the referred prototype. The prototype is composed of the following components:
\begin{itemize}
    \item Data Provider: data is provided by external systems perceived as sensor data providers via the source broker. These data providers have different sensor hubs, i.e., a system that combines many sensor instruments and has at least a communication device that allows the transmission of collected data upstream to the cloud via an MQTT publisher client that functions as a gateway.
    \item Source Broker: MQTT broker responsible for managing raw sensor data streams. There is one instance for each sensor data provider.
    \item Transformer: bridge component implemented on the client side that converts the incoming sensor data from the source broker. The outputted data by the deployed bridge transformation instances reduces the original sizes of sensor data as a result of merging the heterogeneous formats and converting them into a unifying data model.
    \item Destination Broker: MQTT broker that distributes data to the different applications needing data (see application layer in the IoT layered architecture introduced in Section~\ref{sec:background}).
    \item Data Consumer: MQTT subscriber client that functions as an entry point of real-time sensor data in the application for which the system was designed.
    \item Forwarder: MQTT subscriber client that forwards data using different application layer protocols, e.g., REST (Excluded in our study).
    \item Data Warehouse: data space service responsible for storing incoming sensor data and metadata and providing them as historical data (Excluded in our study).
    \item Monitoring Subsystem: stores data collected by the Monitoring Agent in the IoT Data Platform. 
\end{itemize}
We have used the HiveMQ community edition as the broker, which is available open-source\footnote{\url{https://www.hivemq.com/community/open-source/}}.
For the benchmark, only the Data Consumer MQTT subscriber was deployed for the study while we excluded both the Forwarder and Data Warehouse components.
Our main focus is the data delivery to the application layer via MQTT. 

Concerning the configurations for the MQTT clients (publishers represented by the gateways and the subscriber), we configured 60 seconds for the connection's KEEP\_ALIVE flag. We also used the MQTT CLEAN\_START flag when connecting to the broker to ensure no persistency between each of the 10 repetitions ran. Lastly, concerning the session persistence during each execution run, on the publisher side, we emulated the real-world behavior in which the MQTT clients disconnect after successfully sending the message and then connecting again to send the next one. This happens because the gateways publish data to the source brokers using mobile or satellite networks. On the other hand, since the subscriber is within the same operational boundary as the destination broker and within the same infrastructure and network, the connection is made at the beginning of each execution run. It is torn down when the execution is halted.

We used the MQTT-related data from the brokers~\cite{hivemq_metrics} recorded in the IoT platform’s monitoring subsystem during the benchmark execution. We used the docker stats command to collect resource usage from both host machines. In host machine 2, we verified the resource usage data from the monitoring system for the measurement validation. Both data recorded are available in the replication package.
For the latency metric, the end-to-end delay was measured by subtracting the reception time of the messages from the transmission time. The  transmission time was generated as a user property information sent alongside the published message. The reception time was generated at the subscriber component upon reception of messages but also records the arrival time, topic name, order, and ID of the incoming message. The ID was used to calculate the number of duplicated messages which could happen for the executions of the benchmark for QoS level 1 of the deployment options.
The recording of the values were done at the subscriber component which logged them into files (execution logs with the recorded raw measurements which are available in the replication package). Execution logs with MQTT messages transmission and their sizes were also recorded during the experiments for the load drivers and the transformer bridge components. 

For the bridge architecture under test components, we allocated a limit of 512 MB of memory and half CPU core for each deployed component in option 1. For option 2, the same total amount was equally divided among the 4 deployed components.
In terms of hardware for the benchmark execution, two host virtual machines running on different networks are used as hosts for the benchmark load driver and the deployment of the architectures under test.
Each benchmark component depicted in Figure~\ref{fig:case} ran in an docker container deployed alongside their related components in a docker compose. Table~\ref{tab:specs}, summarizes the host machines resources specifications, capacity, and the associated MQTT components allocated to each machine.

\begin{table}[htb!]
\centering
\resizebox{\columnwidth}{!}{%
\begin{tabular}{|l|c|c|c|c|l|}
\hline
\rowcolor[HTML]{9B9B9B} 
\multicolumn{1}{|c|}{\cellcolor[HTML]{9B9B9B}\textbf{Host}} & \textbf{CPU} & \textbf{Mem.} & \textbf{Network} & \textbf{Provisioning} & \multicolumn{1}{c|}{\cellcolor[HTML]{9B9B9B}\textbf{MQTT Component}} \\ \hline
1 & \begin{tabular}[c]{@{}c@{}}8 virtual cores \\ @2.5 GHz\end{tabular} & 2 GB & \begin{tabular}[c]{@{}c@{}}1000 Mb/s\\ Ethernet\end{tabular} & On-premises & \begin{tabular}[c]{@{}l@{}}3 Publishers (Gateways)\\ 1 Subscriber\end{tabular} \\ \hline
2 & \begin{tabular}[c]{@{}c@{}}8 virtual cores \\ @ 2.90GHz\end{tabular} & 16 GB & \begin{tabular}[c]{@{}c@{}}50000 Mb/s \\ Ethernet\end{tabular} & Public Cloud & \begin{tabular}[c]{@{}l@{}}2 Source Brokers\\ 1 Receiving Broker\end{tabular} \\ \hline
\end{tabular}
}
\caption{Benchmark host machines setup configuration.}
\label{tab:specs}
\end{table}

A (virtual) host machine with dedicated resources was allocated to run the source brokers and the architectural options (which includes the destination broker),  alongside the monitoring subsystem. Another (virtual) host machine on servers on-premises was used to run the load driver and the subscriber to ensure the accuracy of the end-to-end delay measurements in the experiments.

We kept the same load model for both benchmark deployment architecture options and allocated the same amount of memory and CPU limit for the bridge architecture under test components. Each benchmark parameter experiment was repeated 10 times, considering the potential effects of random factors when running experiments in the cloud and across different networks.

\subsection{Benchmark Parameters}
\label{sec:bench_params}
Overall we configured different topic names, deployment options of the bridge component, QoS levels, and different payload sizes (small, medium, and large). 
We explore the MQTT topic name subscription in the architectural options configuration to deploy one bridge component per sensor data provider (AUT 1), in contrast with the second option of deploying one bridge component per sensor hub data stream (AUT 2). 
In the production system, the AUT 2 version is used; thus, it is the baseline for defining the deployment and configuration alternatives. AUT 2 has a one-to-one mapping between the number of bridge components deployed and the number of sensor hubs deployed at the remote operational field. In AUT 1, we chose to map the number of sensor data providers to the number of deployed bridge components based on the premise that each sensor data provider has a single data format. Thus, one deployed bridge component can transform data coming from their sensor hubs, resulting in one bridge component and associated link per sensor data provider.
\newline
Consequently, we used the wildcard subscription filter to configure the bridge components for each sensor data provider, being able to capture data from all its sensor hubs by the structure of the topic name where data is published. This configuration option makes it easier to adapt to changes on the sensor hub side over time. 
As an alternative to this configuration, we also investigate in the benchmark whether having the same topic names as AUT 2 impacts the quality attributes mentioned in \textbf{RQ2}. These options imply that each deployed bridge component processes less data in AUT 2 because there is one instance per sensor hub stream. In contrast, AUT 2 has more associated network links than AUT 1.

As previously mentioned, the IoT platform prototype is being fed data from two sensor data providers, whose sensor-based systems are deployed at sea and operate autonomously. Because of the different architecture of the data provider systems, the gateways in Figure~\ref{fig:aut} have different node topology and resulting sensor data sizes, despite both having the same number of sensor hubs deployed.
Overall, a summary of the parameters configured in the two deployment architectures options is listed in Table~\ref{tab:auts}.
\begin{table}[hb!]
\centering
\resizebox{\columnwidth}{!}{%
\begin{tabular}{|l|c|l|l|}
\hline
\rowcolor[HTML]{9B9B9B} 
\multicolumn{1}{|c|}{\cellcolor[HTML]{9B9B9B}Options} & Bridges & \multicolumn{1}{c|}{\cellcolor[HTML]{9B9B9B}Bridge Topics} & \multicolumn{1}{c|}{\cellcolor[HTML]{9B9B9B}\begin{tabular}[c]{@{}c@{}}Bridge Topic \\ Size\end{tabular}} \\ \hline
AUT 1                                                 & 2       & Wildcard subscription per provider                  & 15 bytes                                                                                           \\ \hline
AUT 1                                                 & 2       & List of sensor hubs topics                          & 29 bytes                                                                                           \\ \hline
AUT 2                                                  & 4       & One-to-one mapping to sensor hubs                          & 29 bytes                                                                                           \\ \hline
\end{tabular}%
}
\caption{Deployment architecture configurations.}
\label{tab:auts}
\end{table}

\subsection{Load Configuration}
\label{sec:load}
For the workload generation, we built a load driver for MQTT, since existing tools do not support the payload for each publisher and are not suitable for measuring latency for bridge broker architecture.
Furthermore, considering the system's characteristics described at the beginning of this section, additional requirements apply to the load generator component in the benchmark. Firstly, the asymmetric number of MQTT clients, having many publishers to one subscriber. Secondly, custom MQTT payload in terms of content, but also custom payload definition and transmission per publisher. And lastly, the MQTT topic mapping definition between publishers and subscribers due to the bridging of brokers.

The load follows the fan-in model, which in MQTT translates to having many MQTT publisher client systems (gateways) delivering data to one single MQTT subscriber client. By having sensor hubs that collect data from many sensors for transmission to the gateways, higher volumes of data are generated per transmission hop inside the sensor network. This characteristic of the prototype resulted in a higher volume of data to be used as payload in the published MQTT messages. 

For the first provider, one sensor hub generated around 125 KB of payload and the other 35 KB. For the second, each sensor hub generated around 1,5 KB of data as a payload. In the driver, each gateway publishes the data from its sensor hubs sequentially, and each data provider is run concurrently.
The usage of data from real-world sensors created an asymmetric workload for the platform. 
Each sensor hub payload was published 1000 times by its associated gateway during each execution run and at one batch of sequential transmissions per second (see Table~\ref{tab:load}). Each gateway of 
sensor data provider 1 (top one in Figure~\ref{fig:aut}) published 1000 messages. For sensor provider 2 (bottom one in Figure~\ref{fig:aut}), all the sensor hubs' payloads were published by the only gateway, and thus, it published 2000 messages per execution. The load for each benchmark configuration evaluated contained 4000 messages whose transmission was repeated 10 times.\footnote{The described configurations can be found in the replication package.}
Because of the deployed architecture on the sensor data provider side in the benchmark, the resulting ingestion rate for provider one was one message per second (1 msg/s per gateway), and for provider 2, two messages per second (2 msg/s). An overview of the load driver specification is provided in Table~\ref{tab:load}. 
\begin{table}[h]
\centering
\resizebox{\columnwidth}{!}{%
\begin{tabular}{|c|c|c|c|l|c|}
\hline
\rowcolor[HTML]{9B9B9B} 
Provider                     & Gateway   & \begin{tabular}[c]{@{}c@{}}Sensor \\ Hubs\end{tabular} & \begin{tabular}[c]{@{}c@{}}Load \\ per Hub\end{tabular} & Topic Size & Rate    \\ \hline
                             & Gateway 1 & 1                                                      & 125 KB                                                  & 29 bytes   & 1 msg/s \\ \cline{2-6} 
\multirow{-2}{*}{Provider 1} & Gateway 2 & 1                                                      & 35 KB                                                   & 29 bytes   & 1 msg/s \\ \hline
Provider 2                   & Gateway 1 & 2                                                      & 1,5 KB                                                  & 29 bytes   & 2 msg/s \\ \hline
\end{tabular}%
}
\caption{Benchmark fixed load summary.}
\label{tab:load}
\end{table}
It is important to mention that the topic name where the gateways from each provider published were the same for the different benchmark parameters evaluated.  
However, for the bridge component in the AUT, the topic name is a studied parameter and varies as shown in Table~\ref{tab:auts}.
The higher volume of data originates from IoT applications where, between the perception and the network layer (see Field Operation Site in Figure~\ref{fig:aut}), there is a gateway node working as a sink for data collected by sensors and relaying the data from the edge of the sensor network to the cloud side for further processing and storage. As a result, instead of having each sensor connect directly to the broker, this aggregator component intermediates both ends.

\subsection{Trust Boundaries}
\label{sec:boundaries}
The cross-organizational scenario brings not only a higher volume of data in this evaluation context when compared with other MQTT architecture solutions benchmark studies but also the insertion of trust boundaries.  These boundaries often require implementing security measures known to bring an overhead to latency due to additional message exchange and processing time due to encryption/decryption~\cite{Wolter2010}. The boundaries are inserted because of the integration with sensor data provider systems and within each organization, where data has to cross different networks.

Figure~\ref{fig:aut} depicts these trust boundaries, borrowing the notation from OWASP Threat Modeling~\cite{OWASP}. The boundary inside the sensor network in the link between the sensor hubs and the gateways was omitted for simplicity, as it is outside the scope of the benchmark.
For the benchmark setup, because MQTT works over TCP/IP, we employed TLS/SSL to encrypt the link between the client and the broker systems. When it comes to authentication and authorization, we used the broker solution's available LDAP feature to enforce the access of the gateways, bridge components, and subscribers to the respective brokers and necessary configured topics. 
\section{Results and Discussion}
\label{sec:results}
\begin{table*}[h!]
\centering
\resizebox{0.8\textwidth}{!}{%
\begin{tabular}{|l|cccccc|ccc|}
\hline
\rowcolor[HTML]{9B9B9B} 
\multicolumn{1}{|c|}{\cellcolor[HTML]{9B9B9B}}                         & \multicolumn{6}{c|}{\cellcolor[HTML]{9B9B9B} AUT~1}                                                                                                                                                                                                                             & \multicolumn{3}{c|}{\cellcolor[HTML]{9B9B9B} AUT~2}                                                              \\ \cline{2-10} 
\rowcolor[HTML]{C0C0C0} 
\multicolumn{1}{|c|}{\cellcolor[HTML]{9B9B9B}}                         & \multicolumn{3}{c|}{\cellcolor[HTML]{C0C0C0}15~bytes  Topic Size}                                                                                             & \multicolumn{3}{c|}{\cellcolor[HTML]{C0C0C0}29~bytes  Topic Size}                                                & \multicolumn{3}{c|}{\cellcolor[HTML]{C0C0C0}29~bytes  Topic Size}                                                \\ \cline{2-10} 
\rowcolor[HTML]{E6E5E5} 
\multicolumn{1}{|c|}{\multirow{-3}{*}{\cellcolor[HTML]{9B9B9B}Metric}} & \multicolumn{1}{c|}{\cellcolor[HTML]{E6E5E5}QoS~0 } & \multicolumn{1}{c|}{\cellcolor[HTML]{E6E5E5}QoS~1} & \multicolumn{1}{c|}{\cellcolor[HTML]{E6E5E5}QoS~2} & \multicolumn{1}{c|}{\cellcolor[HTML]{E6E5E5}QoS~0 } & \multicolumn{1}{c|}{\cellcolor[HTML]{E6E5E5}QoS~1} & QoS~2 & \multicolumn{1}{c|}{\cellcolor[HTML]{E6E5E5}QoS~0 } & \multicolumn{1}{c|}{\cellcolor[HTML]{E6E5E5}QoS~1} & QoS~2 \\ \hline
Latency (ms)                                                           & \multicolumn{1}{c|}{132.3}                         & \multicolumn{1}{c|}{185.3}                         & \multicolumn{1}{c|}{216.9}                         & \multicolumn{1}{c|}{132.9}                         & \multicolumn{1}{c|}{186.7}                         & 217.7 & \multicolumn{1}{c|}{135.6}                         & \multicolumn{1}{c|}{191.2}                         & 221.6 \\ \hline
Published Messages (Gateways)                                                   & \multicolumn{1}{c|}{4000}                          & \multicolumn{1}{c|}{4000}                          & \multicolumn{1}{c|}{4000}                          & \multicolumn{1}{c|}{4000}                          & \multicolumn{1}{c|}{4000}                          & 4000  & \multicolumn{1}{c|}{4000}                          & \multicolumn{1}{c|}{4000}                          & 4000  \\ \hline
Received Messages (End-to-end)                                                    & \multicolumn{1}{c|}{2004}                          & \multicolumn{1}{c|}{4000}                          & \multicolumn{1}{c|}{4000}                          & \multicolumn{1}{c|}{2002}                          & \multicolumn{1}{c|}{4000}                          & 4000  & \multicolumn{1}{c|}{2004}                          & \multicolumn{1}{c|}{4000}                          & 4000  \\ \hline
Received Messages (Source Brokers)                                                   & \multicolumn{1}{c|}{2005}                          & \multicolumn{1}{c|}{4000}                          & \multicolumn{1}{c|}{4000}                          & \multicolumn{1}{c|}{2003}                          & \multicolumn{1}{c|}{4000}                          & 4000  & \multicolumn{1}{c|}{2005}                          & \multicolumn{1}{c|}{4000}                          & 4000  \\ \hline
Received Messages (Dest. Brokers)                                                   & \multicolumn{1}{c|}{2004}                          & \multicolumn{1}{c|}{4000}                          & \multicolumn{1}{c|}{4000}                          & \multicolumn{1}{c|}{2002}                          & \multicolumn{1}{c|}{4000}                          & 4000  & \multicolumn{1}{c|}{2005}                          & \multicolumn{1}{c|}{4000}                          & 4000  \\ \hline                       & 4000  \\ \hline
Lost Messages (End-to-end)                                                       & \multicolumn{1}{c|}{1996}                          & \multicolumn{1}{c|}{0}                             & \multicolumn{1}{c|}{0}                             & \multicolumn{1}{c|}{1998}                          & \multicolumn{1}{c|}{0}                             & 0     & \multicolumn{1}{c|}{1996}                          & \multicolumn{1}{c|}{0}                             & 0     \\ \hline
Lost Messages (Source Broker)                                                         & \multicolumn{1}{c|}{1995}                          & \multicolumn{1}{c|}{0}                             & \multicolumn{1}{c|}{0}                             & \multicolumn{1}{c|}{1997}                          & \multicolumn{1}{c|}{0}                             & 0     & \multicolumn{1}{c|}{1995}                          & \multicolumn{1}{c|}{0}                             & 0     \\ \hline
Payload (bytes)                                                        & \multicolumn{1}{c|}{1044}                          & \multicolumn{1}{c|}{5562}                          & \multicolumn{1}{c|}{5562}                          & \multicolumn{1}{c|}{1037}                          & \multicolumn{1}{c|}{5562}                          & 5562  & \multicolumn{1}{c|}{1045}                          & \multicolumn{1}{c|}{5562}                          & 5562  \\ \hline
\end{tabular}%
}
\caption{Benchmark mean results for 10 executions.}
\label{tab:results_global}
\end{table*}

\begin{figure}[] 
	\centering 
    \vspace{-0.3cm}
 \includegraphics[trim={0cm 0cm 0cm 0.8cm},clip,width=\columnwidth]{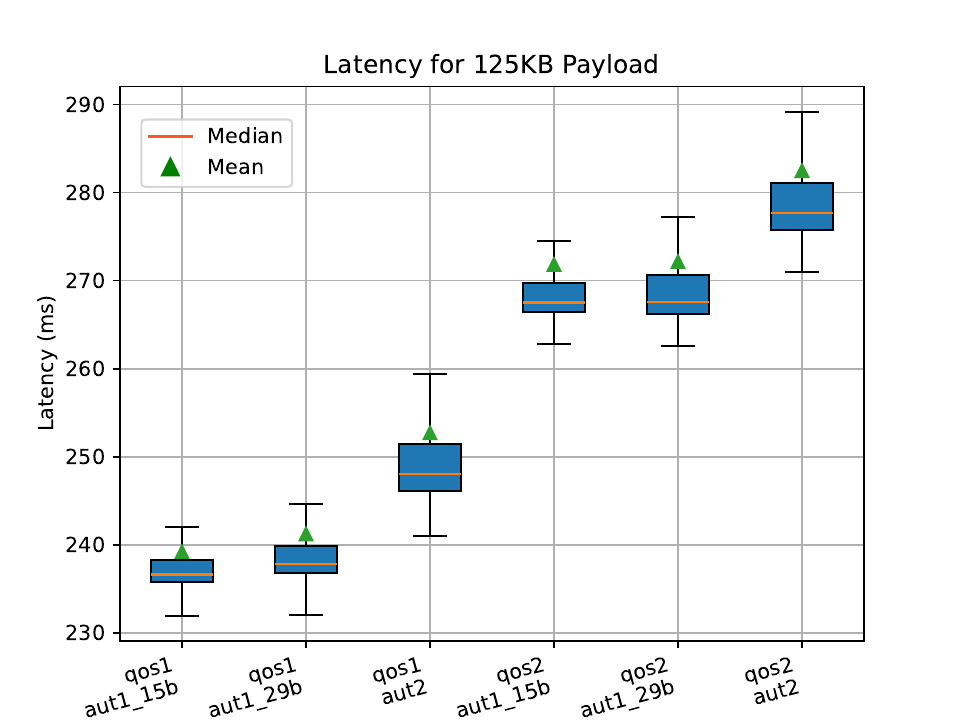}
	\caption{Latency results for Gateway 1 of Provider 1 load with 125 KB payload (see Figure~\ref{fig:aut}).}
    \vspace{-0.3cm}
	\label{fig:results_p1g1}
\end{figure}

\begin{figure}[] 
	\centering 
 \includegraphics[trim={0cm 0cm 0cm 0.8cm},clip,width=\columnwidth]{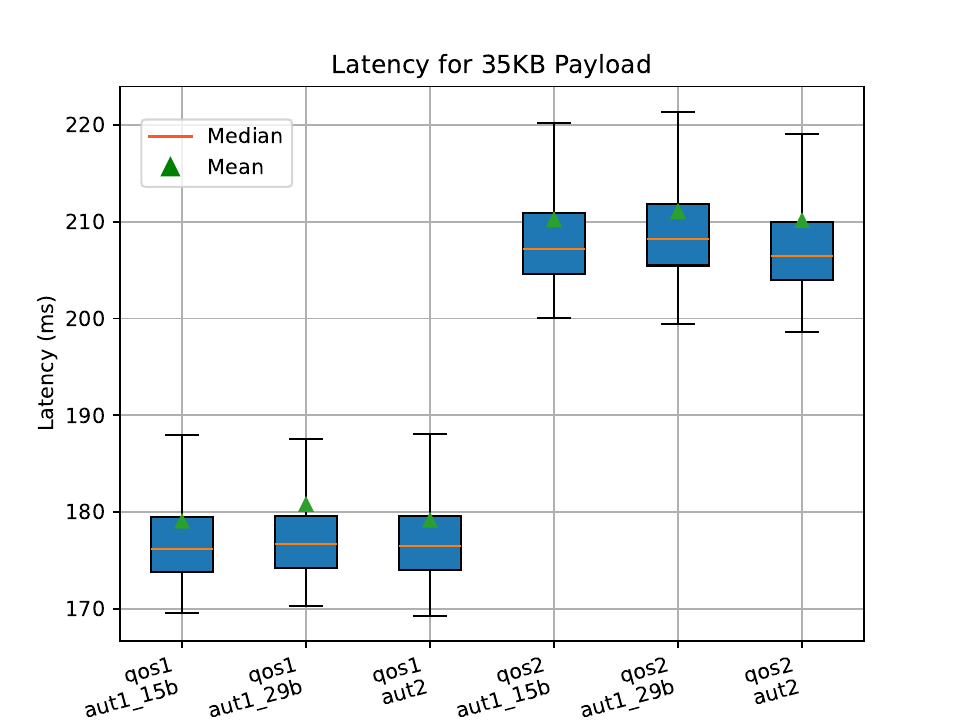}
	\caption{Latency results for Gateway 2 of Provider 1 load with 35 KB payload (see Figure~\ref{fig:aut}).}
	\vspace{-0.5cm}
    \label{fig:results_p1g2}
\end{figure}

\begin{figure}[h!] 
	\centering 
 \includegraphics[trim={0cm 0cm 0cm 0.8cm},clip,width=\columnwidth]{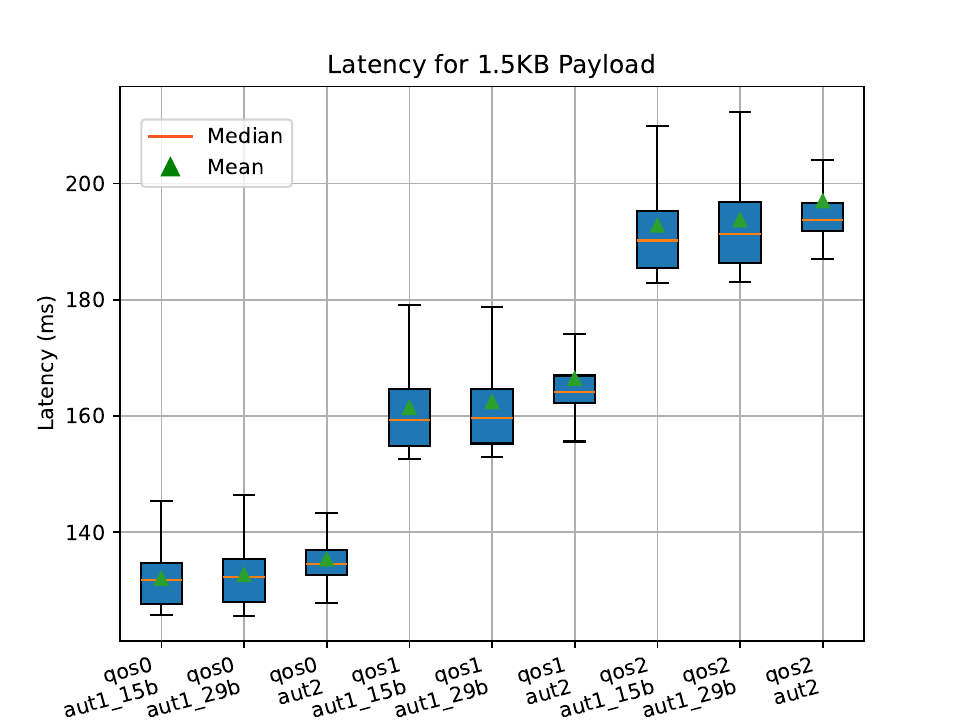}
	\caption{Latency results for Gateway 1 of Provider 2 load with 1.5 KB payload (see Figure~\ref{fig:aut}).}
    \vspace{-0.7cm}
	\label{fig:results_p2g1}
\end{figure}

\begin{figure}[h!] 
\vspace{-0.2cm}
	\centering 
 \includegraphics[trim={0cm 0.3cm 0cm 1.35cm},clip,width=0.7\columnwidth]{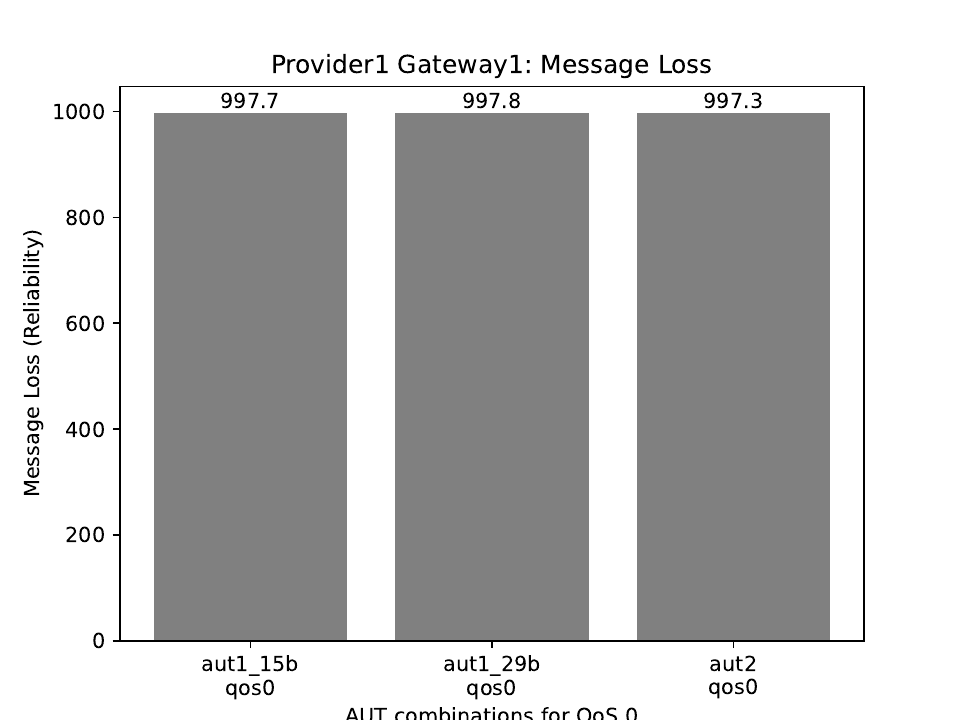}
	\caption{Average message loss for QoS level 0 for Gateway 1 of Provider 1 with 125 KB payload (see Figure~\ref{fig:aut}).}
	\label{fig:results_p1g1_loss}
    \vspace{-0.5cm}
\end{figure}

\begin{figure}[h!] 
	\centering 
 \includegraphics[trim={0cm 0.3cm 0cm 1.35cm},clip,width=0.7\columnwidth]{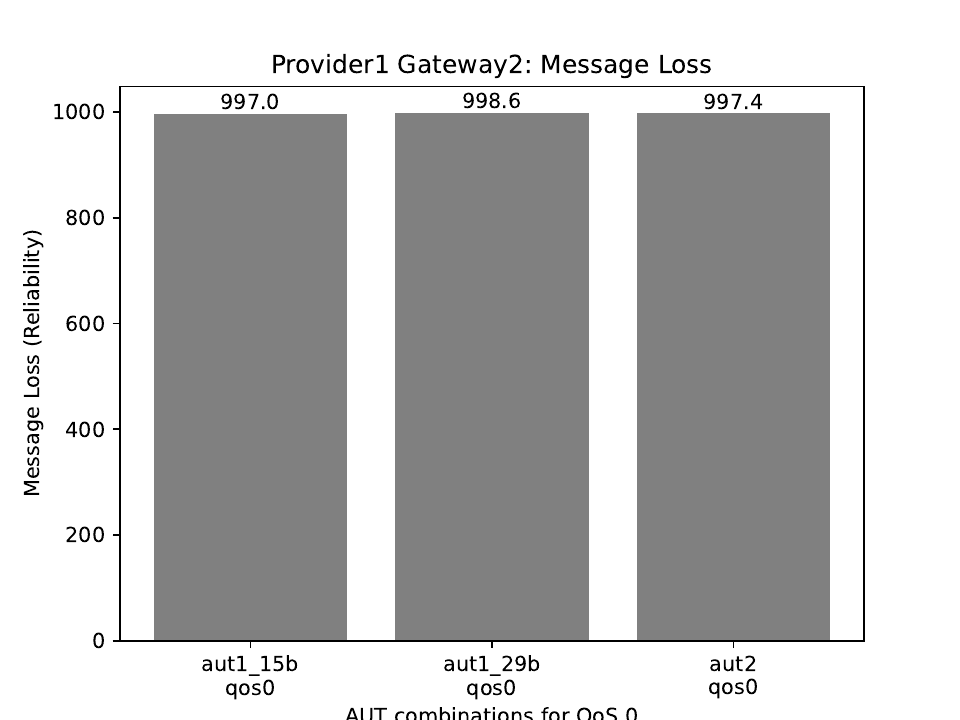}
	\caption{Average message loss for QoS level 0 for Gateway 2 of Provider 1 with 35 KB payload (see Figure~\ref{fig:aut}).}
	\label{fig:results_p1g2_loss}
    \vspace{-0.5cm}
\end{figure}

\begin{figure}[h!] 
	\centering 
    \vspace{-0.1cm}
 \includegraphics[trim={0cm 0.3cm 0cm 1.35cm},clip,width=0.7\columnwidth]{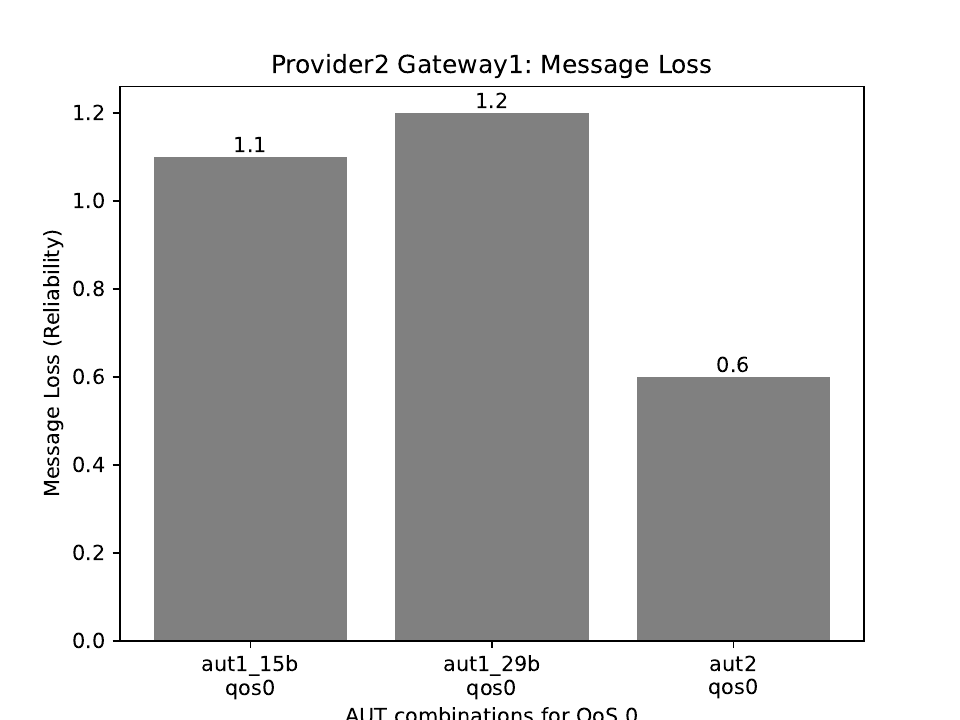}
	\caption{Average message loss for QoS level 0 for Gateway 1 of Provider 2 with 1.5 KB payload (see Figure~\ref{fig:aut}).}
    \label{fig:results_p2g1_loss}
\end{figure}

In this section we provide an overview of the results in Table~\ref{tab:results_global}. This table reflects the aggregated results of each one of the benchmark setups execution which took approximately four hours for the 10 repetitions. The results for the different architectural options and MQTT parameters are presented throughout this section and discussed in the lens of the research questions presented in Section~\ref{sec:rqs}. The raw measurements and monitoring data recorded during the benchmark are available in the replication package, alongside instructions to reproduce the results.  
We start by presenting the average results from the 10 executions for each evaluated setup in Table~\ref{tab:results_global}. The data in the table was aggregated by AUT option and QoS level configured in the benchmark. 
In the table, the average latency for AUT~1 with 29~bytes topic size is very similar to 15~bytes  topic size in AUT~1. However, we notice a marginal difference of a maximum of 1.4 ms in latency. Based on our setup, we assume these differences may be due to the larger topic size of 29 bytes.
Additionally, there is an increasing delay pattern observed  as the QoS level increases in Table~\ref{tab:results_global} for all AUT options. This is consistent with MQTT reliability design as QoS level 0  has the lowest delay and QoS level 2 has the highest delay since more messages are exchanged to recover from failure. 
Furthermore, we analyze the latency and reliability results by payload sizes in Figures~\ref{fig:results_p1g1} to~\ref{fig:results_p2g1_loss}.

Lastly, Figures ~\ref{fig:results_p1g1_loss}, ~\ref{fig:results_p1g2_loss}, and ~\ref{fig:results_p2g1_loss} show the average message loss by provider and gateway filtered by payload size at QoS level 0. For payload size of 125 KB, Figure ~\ref{fig:results_p1g1_loss} shows similar and very high message loss at 997.7 and 997.8 (AUT~1 with 15 bytes and 29 bytes respectively), and 997.3 (AUT~2). For payload size of 35 KB (Figure ~\ref{fig:results_p1g2_loss}), a similar pattern is observed with very high message loss at 997.0 and 998.6 (AUT~1 with 15 bytes and 29 bytes respectively), and 997.4 (AUT~2). For payload sizes of 1.5 KB for sensor provider 2, the message loss is minimal at 1.1 and 1.2 (AUT~1 with 15 bytes and 29 bytes respectively), and 0.6 (AUT~2). In the cases with high message loss, there were not enough data points to calculate average results that could compensate for outliers caused by variability in the real-world network and cloud resources. For this reason, those cases were not considered in the analysis based on payload sizes.

\noindent
\emph{\textbf{RQ1}: How does the deployment architecture of MQTT bridge components impacts message delivery latency and reliability in a cross-organizational data delivery context?}
~\newline
\noindent
The result shows there is a slight difference in delay between the two configurations used for AUT~1 and a more significant difference when comparing both with AUT~2.
These results also apply when the measurements are analyzed and filtered by the provider but do not apply to all payload sizes. The medium payload size (see Figure~\ref{fig:results_p1g2}) presents a smaller difference between the average delays across architectures. The different architectural deployment options do not impact the reliability metrics measured for medium and high payload sizes. The resulting average message losses are very close and consistent for both gateways (see Figures~\ref{fig:results_p1g1_loss}~and~\ref{fig:results_p1g2_loss}). On the other hand, architectural option 2 (AUT~2) performs better for smaller payload sizes, having almost half of the lost messages (see Figure~\ref{fig:results_p2g1_loss}).  We observe from the collected data that the loss is spread
more evenly for AUT~1 when compared with data loss in the ingestion link between the gateways and the source brokers.
The smaller data published on the last MQTT link also confirms this result, with less data loss between the bridge transformer and the subscriber components.

Overall, the result shows a trade-off between the latency and reliability at all QoS levels for small payload sizes, while for medium and large payload sizes, this trade-off only exists for the higher QoS levels (1 and 2).
On average, introducing more bridge processors seems to increase the end-to-end delay from the producer to the subscriber. However, there is no noticeable difference in message losses. We can infer that the transmission cost is higher than the processing cost and recommend that fewer bridge processors be used when composing the bridge architecture.

\noindent
\emph{\textbf{RQ2}: How does the topic name and the payload size impact both latency and reliability of the deployment architecture of MQTT bridge components?}
~\newline
\noindent
   
Only the QoS level 0 presents losses of messages. These losses are significant for the medium (35 KB) and high (125 KB) payload sizes, despite the payload sizes being much smaller than the 256 MB overall packet size limit specified in the MQTT protocol.
However, the results for the small payload size (1.5 KB) for QoS level 0 is much more reliable in terms of the message loss metric.
In Roy et al.~\cite{GUHAROY2018300}, a similar pattern is reported showing a correlation between the payload size and the number of lost messages.
Because both source brokers and the destination broker in the bridge architecture were deployed in the same host machine (see Figure~\ref{fig:case}), the message loss measured mostly affected the link between the gateways and the source brokers (see Table~\ref{tab:results_global}) for the higher payload sizes (see Figures~\ref{fig:results_p1g1_loss} to~~\ref{fig:results_p1g2_loss}).
On average, there is one message loss shared among the links between the bridge and the subscriber components to the destination broker.
In the real system, there could be potentially more losses if both brokers bridged are in different networks.

Regarding the topic name overhead in the delay, from the data in Figures~\ref{fig:results_p1g1} to ~\ref{fig:results_p2g1}, and Table~\ref{tab:results_global}, we can note a slight difference in delay across all payload sizes within the same AUT. In these cases, the 15 bytes topic size consistently presents less average delays compared to the 29 bytes topic size. We ignore cases with significant data loss which occurred with QoS level 0 for medium (35 KB) and large (125 KB) payload sizes.
To adequately draw conclusion of the impact of topic name overhead, more detailed experiments that consider different topic names and sizes will be necessary.

In summary, as a recommendation to practice, QoS level 1 performs well regarding reliability for all payload and topic overhead sizes. This quality-of-service level should be considered if smaller delays are required and mechanisms are implemented to detect duplicated message delivery. Furthermore, the bridge architecture becomes much more unreliable for medium and high payload sizes when operating in QoS level 0 with the loss of almost all transmitted messages. Therefore, this level should be avoided when transmitting with medium and high payloads. In our case, 35 KB and 125 KB. However, QoS level 0 can be considered for small payload (e.g., 1.5 KB in our case) in use cases that can tolerate few message losses and where there is bandwidth constraint. As QoS level 0 has less communication delay. 

Finally, as recommendations for research, MQTT implementations should be investigated regarding different payload size magnitudes, from 1 byte up to the maximum allowed packet size of 256 MB. As our result has shown, this knowledge can further help designers composing MQTT bridge architecture to efficiently decide on trade-offs between latency and reliability for their use cases. In addition, network locality should be considered when conducting such studies to introduce realistic operational scenarios and enable a better assessment of transmission link loss and delay. Lastly, alternative setups can be done by emulating challenging network conditions, including the injection of packet loss.

\section{Threats to Validity}
\label{sec:threats}
Since we applied a benchmark methodology to evaluate different architectural deployment options, it is important to discuss at least the study's construct validity and reliability~\cite{benchmark_std}.

Firstly, concerning construct validity, we used the end-to-end delay as a metric for latency. Considering the MQTT clients, this metric was calculated by subtracting the timestamp added at the publishing point from the timestamp at the subscribing client at the destination bridge. To have consistency and high accuracy when computing the metric, we deployed the clients at both ends in the same host machine to ensure the usage of synchronized clocks, ensuring the validity of the measurements. Furthermore, the primary purpose of the MQTT protocol is to deliver data in IoT contexts, and thus we used message loss as an indicator to evaluate the reliability of the system. By definition in the MQTT protocol, quality-of-service level 0 is the only level where message loss is expected and thus, we conduct the analysis of the reliability for that benchmark parameter. To capture message loss impact, the client and bridge broker components ran on different machines on separated networks, which is a more realistic setup. This is especially true considering that we used QoS level 0 as a parameter, which makes less sense to measure when both clients and brokers are on the same machine or even network.  

Regarding the reliability of the benchmark execution, to minimize the impact of the variability of using cloud infrastructures and different networks, the systems under test were executed under the same resource limits with dedicated resources while repeating each experiment 10 times. Furthermore, raw measurements were also made available in the replication package of this study for more details on how to reproduce the aggregated results presented in Section~\ref{sec:results}.

As regards internal validity, the consistent average payload and topic sizes through the different benchmark setups validate our fixed message load for the replicated sensor data and, thus, the correct and expected behavior of the systems under test. In addition, the HiveMQ broker metrics stored in Prometheus were used to validate MQTT-related measurements done on the subscriber side.

Lastly, our experiment has considered only three payload sizes, two topic names, two deployment options, and one subscriber. We cannot generalize to other configuration setups, for example, when cluster setup is used. More experiments suggested in the recommendation for research will be necessary to generalize.

\section{Related Work}
\label{sec:related-work}
As mentioned in~\cite{Pop2020}, the benchmark definitions for MQTT solutions focus on understanding the Service Level Agreements (SLA) that commercial vendors of such solutions specify. In this study, our setup included the evaluation of MQTT brokers platforms not in isolation, but in context considering the whole data ingestion processing pipeline in a bridge architecture scenario.
Many recent studies have focused on benchmarking metrics such as latency and scalability of existing MQTT broker platforms, evaluating parameters such as the number of broker nodes and clients or QoS level~\cite{gruener21,Koziolek2020,Longo2022}.

Contrary to previous studies on MQTT-related benchmarks, we introduce topic size overhead as a benchmark parameter and use real sensor data, whose volume is in kilobytes. This contrasts with previous studies where the payload is smaller than 100 bytes as the load originates directly from a sensor and does not have a gateway that aggregates data from different sensor hubs.     

Concerning network locality, i.e., the network where MQTT clients and brokers are deployed, some studies have considered this as a benchmark parameter for the MQTT-related benchmark~\cite{Longo2022}. As an alternative. Package loss has also been simulated by injection of failures when both client and brokers are deployed in the same network~\cite{gruener21}. A more sophisticated approach by emulating the over-the-air satellite link between the gateway and the broker was done in~\cite{Soua2022} but QoS levels were not part of the studied benchmark parameters. In our study, we separated the clients from the brokers in our benchmark setup, deploying them on separate host machines across different networks. This setup allowed the assessment of QoS level 0 under realistic network conditions while capturing message losses. 

We do not consider cluster architectures for the logical broker deployment because scalability and high throughput are not concerns/requirements for the deployed system under test, in contrast with latency and reliability.
In~\cite{Koziolek2020} an evaluation of several options with a focus on scalability in cluster setups was performed for different MQTT brokers vendors and EMQX had the lowest average latency for high throughput loads.
Lastly, other alternatives to MQTT, such as CoAP, HTTP, or AMQP, can also be employed in the IoT network layer, depending on the requirements, and have been extensively investigated and compared in~\cite{9615332,kokkonis18,iot_gateways19,mishra2020use,naik17}.

\section{Conclusion and Future Work}
\label{sec:conclusion}
We found that smaller packet size at QoS level 0 can be transmitted reliably, whereas medium and larger packet sizes suffer high message losses at the same level. Reliability and latency conflict for both QoS levels 1 and 2 for all payload sizes. We confirm that MQTT payload size significantly impacts the loss of messages when transmitting with the lower quality of service (QoS level 0). 
In addition, for small and large payload sizes, there is a clear difference between architectural option 1, which has one bridge component per data provider, and architectural option 2, with one bridge component per sensor hub stream. The first option performs better in latency, with smaller average delays. We can highlight that more bridge processors between the source and destination brokers correspond to higher delay. In contrast, fewer processors with less communication links produce less average delays.

As previously mentioned, the data provider architecture should be investigated in future work, particularly the gateway deployment options. The energy efficiency of data transmission concerning different payload sizes could also be considered as a quality attribute.

\providecommand{\doi}[1]{DOI: \href{https://doi.org/#1}{#1}}

\end{document}